\pdfoutput=1

\documentclass[10pt,conference]{IEEEtran}
\IEEEoverridecommandlockouts

\usepackage{listings}
\usepackage{color}
\usepackage{makecell}
\usepackage{float}
\usepackage{longtable}
\usepackage{array}
\usepackage{booktabs}
\usepackage{graphicx}
\usepackage{amsmath}
\usepackage{tcolorbox}
\usepackage{lipsum} 
\usepackage{graphicx}
\usepackage{caption}

\usepackage{threeparttable}  
\usepackage{booktabs}
\usepackage{multirow}
\usepackage{bm}

\definecolor{dkgreen}{rgb}{0,0.6,0}
\definecolor{gray}{rgb}{0.5,0.5,0.5}
\definecolor{mauve}{rgb}{0.58,0,0.82}

\tcbuselibrary{skins, breakable}
\newtcolorbox{chatbubble}[1][]{
    enhanced,
    breakable,
    colback=gray!5,  
    colframe=gray!60, 
    fonttitle=\bfseries,
    boxrule=0.5mm,    
    width=\textwidth, 
    arc=3.5mm,          
    boxsep=1mm,           
    left=0.5mm,             
    right=0.5mm,            
    top=1mm,              
    bottom=1mm ,          
    title=#1,   
    overlay={ 
        \fill[gray!60]
          (frame.south east) ++(0.3mm,0.5mm) circle (1.2mm); 
    }
}

\newcounter{chatlisting}
\renewcommand{\thechatlisting}{Chat Listing \arabic{chatlisting}}

\newenvironment{chatgroup}[1]{%
    \vspace{2mm}
    \refstepcounter{chatlisting} 
    \def\chatcaption{#1} 
    \begin{minipage}{0.95\columnwidth} 
}{%
    \end{minipage}
    \vspace{0mm}\captionsetup{type=figure}\caption*{\thechatlisting: \chatcaption} 
    \vspace{4mm}
}

\AtBeginDocument{%
  }

\begin{document}

\title{Analysis of Student-LLM Interaction in a \\Software Engineering Project}

\author{%
  \IEEEauthorblockN{Agrawal Naman, Ridwan Shariffdeen, Guanlin Wang, Sanka Rasnayaka, Ganesh Neelakanta Iyer}
  \IEEEauthorblockA{School of Computing, National University of Singapore\\
    \{naman.a@u.nus.edu, ridwan@comp.nus.edu.sg, wguanlin@u.nus.edu, sanka@nus.edu.sg, gni@nus.edu.sg\}}
}

\maketitle

\begin{abstract}
Large Language Models (LLMs) are becoming increasingly competent across various domains, educators are showing a growing interest in integrating these LLMs into the learning process. Especially in software engineering, LLMs have demonstrated qualitatively better capabilities in code summarization, code generation, and debugging. Despite various research on LLMs for software engineering tasks in practice, limited research captures the benefits of LLMs for pedagogical advancements and their impact on the student learning process. To this extent, we analyze 126 undergraduate students' interaction with an AI assistant during a 13-week semester to understand the benefits of AI for software engineering learning. We analyze the conversations, code generated, code utilized, and the human intervention levels to integrate the code into the code base.

Our findings suggest that students prefer ChatGPT over CoPilot. Our analysis also finds that ChatGPT generates responses with lower computational complexity compared to CoPilot. Furthermore, conversational-based interaction helps improve the quality of the code generated compared to auto-generated code. Early adoption of LLMs in software engineering is crucial to remain competitive in the rapidly developing landscape. Hence, the next generation of software engineers must acquire the necessary skills to interact with AI to improve productivity.

\end{abstract}

\section{Introduction}

Generative large language models (LLMs) have become crucial in education, excelling in tasks from math problem-solving~\cite{Enkelejda2023} to dialog-based tutoring~\cite{Park2024} and aiding software engineering projects~\cite{Rasnayaka2024}. Their versatility has made them highly sought after in educational settings. In software engineering, LLMs particularly excel in tasks like code summarization~\cite{Ahmed2023}, test generation~\cite{Chen2024}, program analysis~\cite{Zhang2024}, code review~\cite{Lu2023}, bug fixing~\cite{Jin2023}, and code generation~\cite{Bairi2024}. Despite growing interest in AI for education, research remains limited on how students use LLMs for open-ended tasks in software engineering projects.


In this work we examine the interaction between undergraduate students and AI assistants in a software engineering course. Students were tasked with using AI to develop a Static Program Analyzer (SPA) for a custom programming language. Over a 13-week semester, teams of six students undertook various tasks, from requirement engineering to user acceptance testing. They received unlimited premium access to Microsoft CoPilot and OpenAI ChatGPT. At semester's end, we collected all AI-driven conversations, code, and artifacts, along with student-annotated code metadata, for analysis. We examine the collected data to answer the following research questions:

\begin{itemize}
     \item Is there a significant difference between code generated by ChatGPT and CoPilot? $\rightarrow$ We compare code complexities using various metrics.
    \item How does the code evolve during a conversation between a student and AI? $\rightarrow$ We analyze conversation logs and extract code for each conversation. 
    \item What is the impact of using AI assistant on their learning outcomes? $\rightarrow$ We analyze the conversation volume, final code output, and evolution of the prompting techniques. 
    \item Does the interaction between the student and AI result in a positive engagement? $\rightarrow$ We perform sentimental analysis across each conversation. 
\end{itemize}

A total of 126 undergraduate students in 21 groups, generated 730 code snippets (172 tests and 558 functionality implementations) using CoPilot and ChatGPT. We also collected 62 ChatGPT conversations that generated code, amounting to 318 messages between students and ChatGPT. Of the total 582,117 lines of code across all teams, 40,482 lines of code (6.95\%) were produced with an LLM's help. 


Upon analysis, Copilot-generated code is longer and more complex (i.e. higher Halstead Complexity) than ChatGPT's, making it harder to interpret. Despite initial assumptions, student feedback shows no significant difference in the integration effort required for both Copilot and ChatGPT-generated code. Furthre analyzing the conversation logs, we identified that through feedback ChatGPT generated code meets project needs with minimal refinement. Sentiment analysis of the conversation reveals on average the conversation ends on a positive note. Indicating conversational-based assistance generate code requiring minimal manual refinement. Over the semester, we also observed a noticeable improvement in the quality of the prompts generated by students, demonstrating their growing ability to craft more effective and precise prompts for better outcomes.

Based on the observations from our study, we discuss design considerations for a future educational course tailored to using AI assistants for software engineering. These considerations include promoting students to learn better prompting strategies and evolving the use of AI assistants beyond merely being a tool for code generation. Our contribution lies in providing an in-depth analysis of how students use ChatGPT in a project-based software engineering course.  
\section{Methodology}\label{sec-case}

\textbf{Project Description:}
In compliance with institutional guidelines, approval for our research was obtained from the Departmental Ethics Review Committee (DERC) before conducting the study. The undergraduate level software engineering course within which this study is conducted involves a 13-week long, robust software development project, where 126 students are tasked with building a Static Program Analyzer (SPA), with three distinct milestones (MS1, MS2, MS3) where the delivery of a functional SPA is expected. The SPA is capable of performing analysis on a course specific custom programming language. The structure of the project is similar to the project used in \cite{Rasnayaka2024}, where the SPA is further subdivided into:
\begin{itemize}
    \item A Source Parser (SP) which analyzes the custom language to extract abstractions such as design entities.
    \item A Program Knowledge Base (PKB), responsible for storing the extracted information.
    \item A Query Processing Subsystem (QPS), which is able to handle queries written in an SQL-like language for querying the PKB, and provide responses to the user.
\end{itemize}

Throughout the development phase, students were granted organizational access to the paid versions of ChatGPT via both the ``Chat'' and ``Playground'' interfaces, enabling close monitoring of their usage. Additionally, students were also able to access  GitHub Copilot features through their institutional GitHub Pro accounts. Access to both of these LLM code generators is funded by the university. Students are also actively encouraged to utilize LLMs and integrate them into their development cycle, and the usage of their organizational access was also reserved strictly for the purposes of this project. Through this setup, we are able to obtain data regarding students' interactions with LLMs, and also the conversational history and information about prompts that were used on ChatGPT.

\textbf{Code extraction and ChatGPT conversations:} Following our initial work~\cite{Rasnayaka2024} we extracted LLM-generated code snippets used by the students at each milestone, which was achieved by requiring students to tag the LLM-generated code utilized in their project with the following information:

\begin{itemize}
    \item Generator used to obtain the output code.
    \item Level of human intervention required to modify the code.
    \item Link to the conversation (only for ChatGPT)
\end{itemize}

The tagging and collection of student data, as well as the definitions of human intervention levels (0, 1, and 2), follow our previous work in \cite{Rasnayaka2024}: level 0 (no changes), level 1 (10\% or fewer lines changed), and level 2 (more than 10\% of the lines changed). This paper introduces a new aspect by including links to student-LLM conversations.


The collected data at each milestone is cumulative, reflecting students' iterative development of their SPA over the semester. We also gathered data on students' use of ChatGPT, including the prompts and generated code, to analyze how conversational interactions affect the quality and usability of LLM-generated code.

\textbf{Overview on Analysis:}
This study explores how students in a software engineering course interact with LLMs like ChatGPT and GitHub Copilot, to use the LLM generator tools to help them in their development process, particularly in the context of how approaches to code generation and the generated outputs evolve. By examining how students interact with these LLMs, adapt generated code, and refine their prompting strategies, we aim to reveal the dynamics of human-AI collaboration in SE education. Here we unpack a multi-layered analysis that spans the quality and complexity of LLM-generated code, the processes of integrating LLM-generated code within student repositories, and the evolving conversational interactions between students and LLMs across milestones.

To comprehensively analyze the quality of LLM-generated code, we used a set of four distinct metrics: total lines of code (LOC), cyclomatic complexity, maximum control flow graph (CFG) depth, and the Halstead effort metric. These metrics provide insights into the sophistication and structural intricacies of the code produced by LLMs. 

\begin{itemize} 
\item \textbf{Total Lines of Code (LOC)}: Serves as a basic indicator of code verbosity and has been used to estimate the programming productivity of a developer.

\item \textbf{Cyclomatic Complexity}: Measures the number of linearly independent paths within the code, and evaluates its logical complexity. Higher cyclomatic complexity can be indicative of maintainability challenges.

\item \textbf{Maximum Control Flow Graph (CFG) Depth}: Measures the depth of nested structures within the code. Increased CFG depth can reflect the presence of deeply nested loops or conditional statements, which may complicate code comprehension and maintenance.

\item \textbf{Halstead Effort}: Estimates the mental effort required to understand and modify the generated code. Higher values suggest that the code may be more challenging to understand and maintain.
\end{itemize}

This work extends our previous research \cite{Rasnayaka2024} by adding a new dimension of sentiment analysis enabled by the collected prompts, providing insights into student - AI interactions. We also introduced new metrics, offering deeper analysis of how code quality and usability vary across different generation approaches.

\section{Results}\label{secres}


\subsection{Analysis of LLM Usage}
We first analyzed the code snippets generated using LLMs across each milestone for each team. Table~\ref{tab:complete_teams_model_usage} captures the cumulative model usage within each team. 5 teams did not use any LLMs for code generation tasks despite providing premium access for the project. 
Out of the remaining 16 teams, 12 used LLMs to generate a moderate number ($>$10) of code snippets. Among these, 6 primarily relied on Copilot, 5 heavily utilized ChatGPT, and 1 team used both tools equally. 

Analyzing across milestones significant decline can be observed in usage of both ChatGPT and Copilot by all teams. This suggests the student teams heavily on AI assistants to generate code earlier in the course but reduced in the latter stages of the course. For some of the teams, we observe a decline in the cumulative number of code snippets from the first to the last milestone. Notably, teams 5 and 8 generated fewer Copilot snippets in the third milestone compared to the second. A similar trend is evident for ChatGPT-generated code in teams 10, 13, and 17. This suggests that some AI-generated code from earlier milestones was either refactored or removed entirely by the end of the project.



\begin{table}[t!]
\centering
\scriptsize
\caption{Cumulative model usage (code snippets generated \& accepted into codebase) per team across ChatGPT and Copilot}
\label{tab:complete_teams_model_usage}
\resizebox{0.3\textwidth}{!}{%
\begin{tabular}{r *{6}{c}}
\toprule
\textbf{TID} & \multicolumn{3}{c}{\textbf{ChatGPT}} & \multicolumn{3}{c}{\textbf{Copilot}} \\
\cmidrule(lr){2-4} \cmidrule(lr){5-7}
 & \textbf{M1} & \textbf{M2} & \textbf{M3} & \textbf{M1} & \textbf{M2} & \textbf{M3} \\
\midrule
1  &    0 &    6 &    9 &    0 &   19 &   20 \\
2  &    3 &    4 &    4 &    0 &    0 &    0 \\
3  &   44 &   44 &   44 &    1 &    1 &    1 \\
5  &   10 &   10 &   20 &    0 &    3 &    2 \\
6  &   19 &   13 &   13 &  164 &  210 &  235 \\
7  &   45 &   54 &   57 &    8 &    9 &    9 \\
8  &    1 &    1 &    1 &   34 &   37 &   27 \\
9  &    7 &   10 &   10 &   22 &    8 &   10 \\
10 &   16 &    9 &    7 &    0 &    0 &    0 \\
12 &    5 &    9 &    9 &   16 &   25 &   27 \\
13 &    6 &    6 &    3 &    0 &    0 &    0 \\
16 &    8 &    9 &   12 &    3 &    3 &    3 \\
17 &   10 &    8 &    6 &    8 &   11 &   10 \\
19 &   14 &   15 &   15 &   79 &  161 &  162 \\
20 &    0 &    0 &    0 &    0 &    1 &    1 \\
21 &   12 &   13 &   13 &    0 &    0 &    0 \\
\midrule
\textbf{Sum} & 200 & 211 & 223 & 335 & 488 & 507 \\
\bottomrule
\end{tabular}%
}
\begin{tablenotes}

\item {
\begin{center}
\textbf{TID}: Team ID, \textbf{M1-3}: Milestone 1-3
\end{center}
}
\end{tablenotes}
\end{table}

\subsection{Analysis of LLM Generated Code}

We analyzed 730 code snippets generated using ChatGPT and Copilot. Table~\ref{tab:code_test_summary} summarizes the types of code snippets produced by both tools, categorized into those for testing purposes and functionality implementation. The analysis shows that students primarily relied on AI assistants for functionality implementation, with moderate usage for generating test cases.

\begin{table}[t!]
\centering
\scriptsize 
\caption{Cumulative model usage (code snippets generated
\& accepted into codebase) per team across test and code generation}
\label{tab:code_test_summary}
\resizebox{0.3\textwidth}{!}{
\begin{tabular}{r *{6}{c}}
\toprule
\textbf{TID} & \multicolumn{3}{c}{\textbf{Test}} & \multicolumn{3}{c}{\textbf{Code}} \\
\cmidrule(lr){2-4} \cmidrule(lr){5-7}
 & \textbf{M1} & \textbf{M2} & \textbf{M3} & \textbf{M1} & \textbf{M2} & \textbf{M3} \\
\midrule
1  &    0 &    6 &    6 &    0 &   19 &   23 \\
2  &    2 &    2 &    2 &    1 &    2 &    2 \\
3  &   11 &   11 &   11 &   34 &   34 &   34 \\
5  &    9 &   10 &   13 &    1 &    3 &    9 \\
6  &   24 &   45 &   53 &  159 &  178 &  195 \\
7  &   33 &   38 &   41 &   20 &   25 &   25 \\
8  &    0 &    0 &    0 &   35 &   38 &   28 \\
9  &   19 &   13 &   13 &   10 &    5 &    7 \\
10 &    0 &    0 &    0 &   16 &    9 &    7 \\
12 &    5 &    9 &   10 &   16 &   25 &   26 \\
13 &    0 &    0 &    0 &    6 &    6 &    3 \\
16 &    3 &    5 &    6 &    8 &    7 &    9 \\
17 &    7 &    9 &    7 &   11 &   10 &    9 \\
19 &    1 &    2 &    2 &   92 &  174 &  175 \\
20 &    0 &    0 &    0 &    0 &    1 &    1 \\
21 &    8 &    8 &    8 &    4 &    5 &    5 \\
\midrule
\textbf{Sum} &  122 &  158 &  172 &  413 &  541 &  558 \\
\bottomrule
\end{tabular}%
}
\begin{tablenotes}
\item {
 \begin{center}
     \textbf{TID}: Team ID, \textbf{M1-3}: Milestone 1-3
 \end{center}

} 

\end{tablenotes}
\end{table}

\begin{figure}[b]
    \centering
    \includegraphics[width=0.9\linewidth]{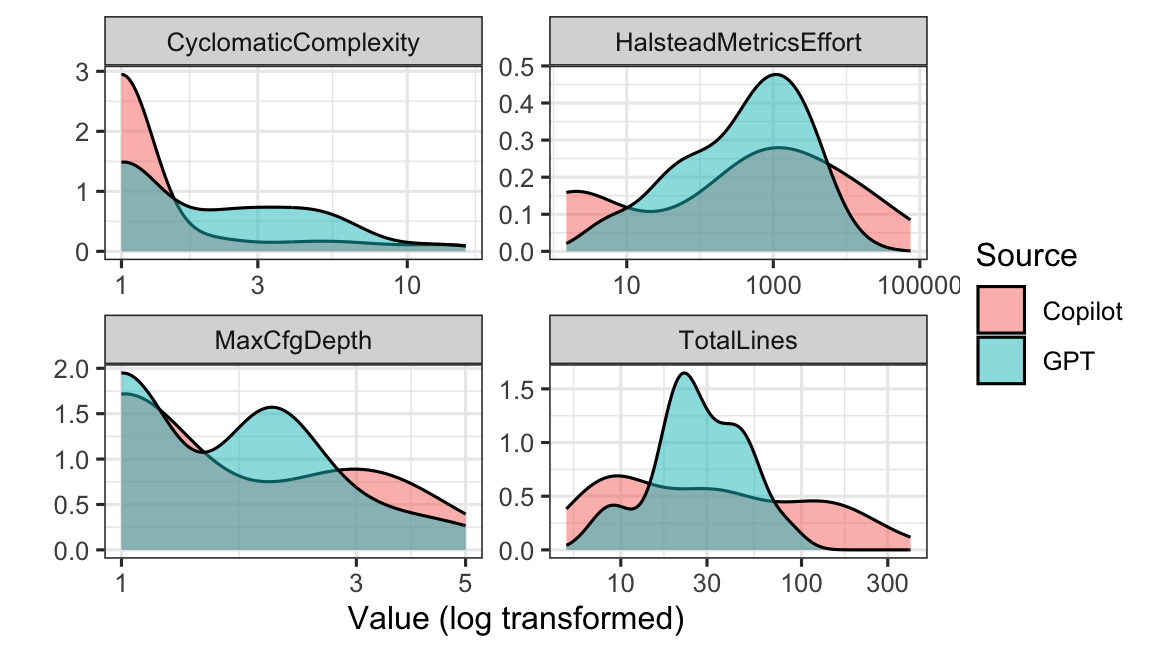}
    \caption{Density Plot for measured key metrics}
    \label{fig:fig1.1.1}
\end{figure}

We further analyzed the complexity of AI-generated code across the three project milestones (MS1, MS2, and MS3) using metrics such as lines of code and cyclomatic complexity. Analysis of AI-generated code revealed a trend towards higher complexity, particularly in code generated by Copilot, as shown by skewed density plots in Figure \ref{fig:fig1.1.1}. This suggests that AI assistance may lead to more complex solutions, although the majority of student-generated code remained moderately complex. Although the average complexity (cyclomatic complexity and total lines) of student-generated code remained moderate, the analysis revealed that AI assistance, particularly Copilot, occasionally produced highly complex solutions, sometimes exceeding the average values by 40 to 50 times. This suggests that AI-generated code, while often effective, has the potential to introduce unnecessary complexity if adopted without careful review and refinement.

Copilot generated significantly more outliers than GPT across all complexity metrics, indicating a tendency toward producing more complex and verbose code. This difference likely stems from Copilot's auto-completion approach, which favors extensive code generation based on common patterns, potentially leading to inflated complexity compared to GPT's more concise and conversationally guided output.

GPT's conversational interface allows for iterative refinement of code, enabling students to guide the model towards simpler and more maintainable solutions. Conversely, Copilot's auto-completion approach, while efficient, can lead to overly complex code due to the lack of nuanced interaction. Additionally, the study's analysis of GPT-generated code is more precise due to the ability to track exact model outputs, while Copilot's contributions are assessed through student modifications, highlighting a difference in how interactions with each tool are measured.

We also analyzed students' efforts to integrate AI-generated code into the project based on reported manual intervention ratings. For Copilot-generated code, the majority (53.6\%) required minor intervention (level 1), while a significant portion (30.0\%) required moderate intervention (level 2), indicating a higher demand for user input to refine or simplify the code. Only 15.2\% of Copilot-generated code required no intervention. In contrast, ChatGPT-generated code more often aligned with user expectations, with 26 \% requiring no manual modification and 22.9\% needing moderate intervention, suggesting ChatGPT-generated code generally met project requirements with minimal refinement.

\begin{figure}[t]
    \centering
    \includegraphics[width=0.7\linewidth]{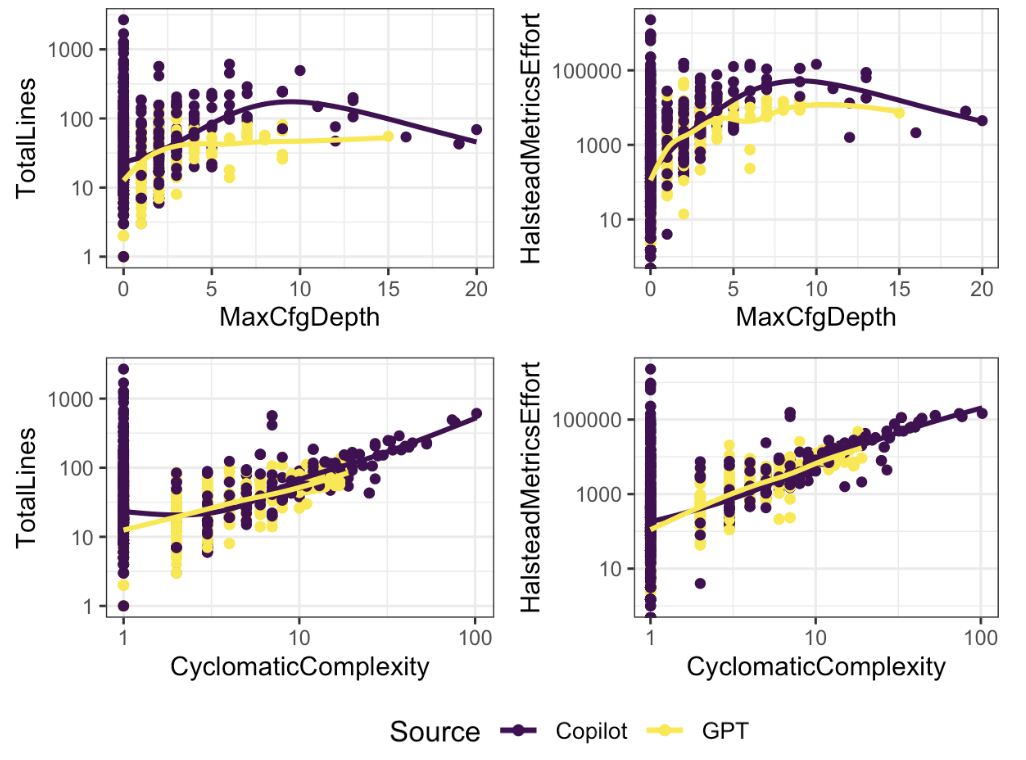}
    \caption{Comparison of ChatGPT and Copilot Complexity Across Various Complexity Measures}
    \label{fig:fig1.1.2}
\end{figure}

We further analyzed the code snippets to understand the difference in complexity between ChatGPT and Copilot-generated code. While GPT and Copilot can achieve similar levels of code complexity, GPT generally does so with less code and lower cognitive effort, as measured by Halstead effort (figure \ref{fig:fig1.1.2}). This suggests that while ChatGPT-generated code shares a similar level of complexity with Copilot's, it is often more concise and easier to understand. ChatGPT's conversational interface enables users to iteratively refine prompts, resulting in more efficient code generation.

\begin{figure}[t]
    \centering
    \includegraphics[width=0.8\linewidth]{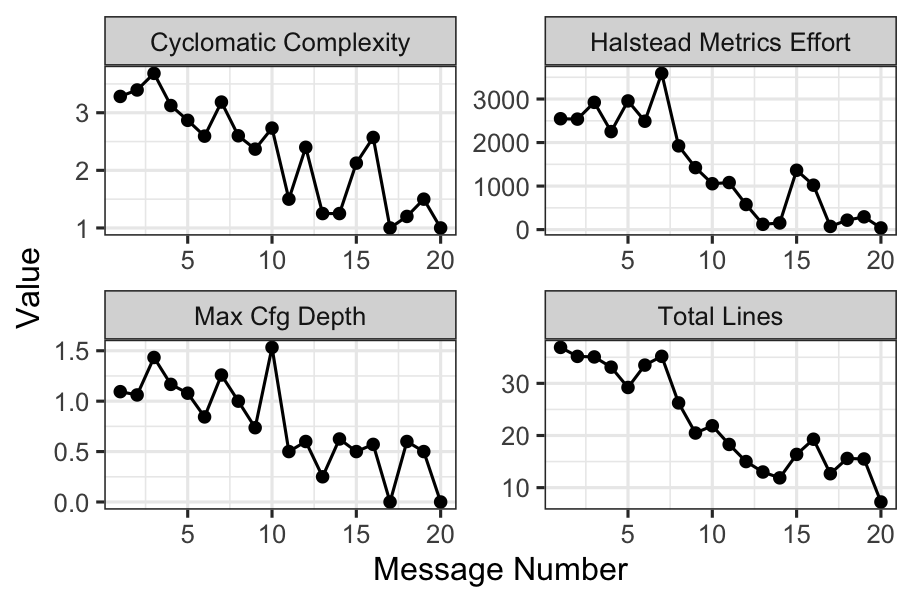}
    \caption{Variation of ChatGPT generated code in a conversation}
    \label{fig:fig1.1.3}
\end{figure}

Analyzing interactions with ChatGPT highlights how this iterative process reduces code complexity. Figure \ref{fig:fig1.1.3} shows the analysis of average code complexity across each conversation. This reveals a consistent trend:  as students converse with ChatGPT, the average complexity of the generated code decreases, particularly in terms of cognitive effort as captured using Halstead Effort. This demonstrates that the interactive nature of ChatGPT allows for iterative refinement and simplification, ultimately supporting the development of more manageable and effective code solutions.  

For example, consider the conversation shown in Chat Listing 1. The student begins by asking ChatGPT to create a C++ function to format an expression string according to specific grammar rules. Over a series of messages, the student requests to simplify the code, asking GPT to “shorten the code” and then to further “abstract into functions if needed.” Each subsequent request leads to a more streamlined and modular version of the code, showing how GPT’s responses become progressively aligned with the student’s preference for conciseness. 

\begin{chatgroup}{A sequence of prompts used by a student to instruct ChatGPT to generate and iteratively improve the generated code.}
    \begin{chatbubble}
     \scriptsize
    Given a string of an expression following these grammar rules: ... Give a function in C++ to convert the string which may not have all of these tokens separated by a whitespace, into a string where all these tokens are separated by a single whitespace.
    \end{chatbubble}
    \begin{chatbubble}
     \scriptsize
    Variable names or constant values can be multi-char.
    \end{chatbubble}
    \begin{chatbubble}
     \scriptsize
    Shorten the code, abstracting it into functions if needed.
    \end{chatbubble}
    \begin{chatbubble}
     \scriptsize
    Shorten the code.
    \end{chatbubble}
    \begin{chatbubble}
     \scriptsize
    Simplify the code for addWhitespace.
    \end{chatbubble}
\end{chatgroup}

This example illustrates how GPT's conversational interface empowers students to iteratively refine code, achieving both reduced complexity and improved maintainability. This adaptability makes GPT well-suited for educational contexts that prioritize clear and concise coding practices. While GitHub Copilot also includes a chat feature, it was not extensively used during the project timeline, as students primarily utilized Copilot for code completion and debugging rather than conversational interactions. Therefore, the study focuses solely on GPT-generated code to further investigate how conversational interactions can enhance code simplicity and efficiency, aligning with the course's objectives.

\subsection{Generated Code vs Integrated Code}
Our next analysis focuses on how students modify and integrate ChatGPT-generated code into their project repositories. Copilot-generated code is excluded, as it lacks the conversational context that evolves the code. This analysis aims to uncover patterns in student adaptations, examining whether they enhance, simplify, or otherwise alter the initial code provided by ChatGPT. We compared each team's repository code with the corresponding ChatGPT-generated code for each conversation. We identified the segment of ChatGPT code with the highest average similarity to the repository and used it as a reference. Our analysis revealed multiple instances of reuse of ChatGPT-generated code snippets in various parts of the team repository. While most students used a generated code once, some used it multiple times, demonstrating its adaptability.

In 95 instances the generated code was used only once, indicating that the majority of students found LLMs useful for generating task-specific code. In 23 instances students reused the generated code twice in the repository and in 5 cases, three times. Additionally, in 3 cases the generated code was reused four times, indicating that the code became a key tool for certain tasks. One notable example of code reuse involved a team who reused a GPT-generated code segment 13 times across different predicates, demonstrating its modularity. The code was adapted for various predicates like UsesPredicate and ParentPredicate, with minor adjustments for logic and parameter types. This highlights the flexibility of the generated code and its strategic use across different parts of the repository.

\begin{figure}[t]
    \centering
    \includegraphics[width=0.8\linewidth]{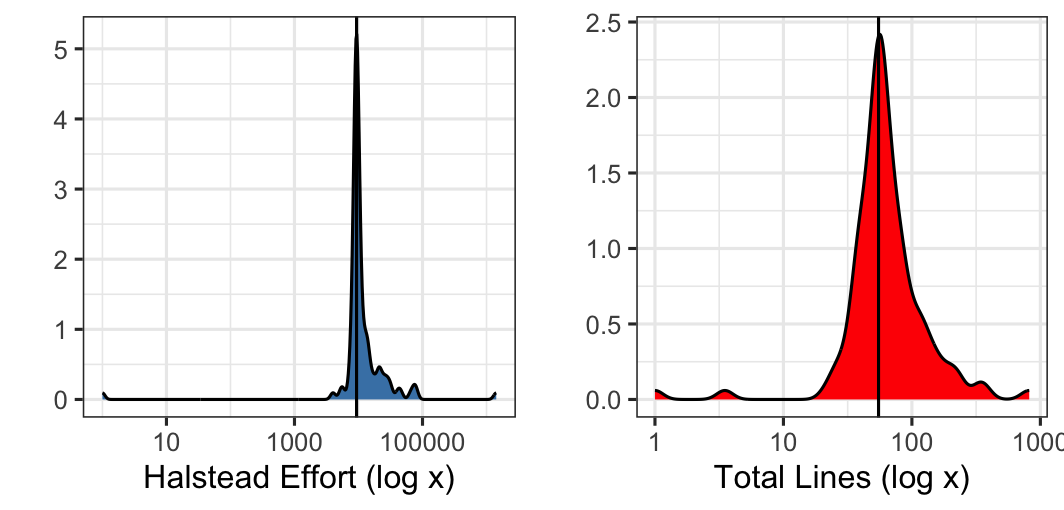}
    \caption{Distribution of Difference Complexity Measures between Repo and GPT Code with Log Transformed x-axis}
    \label{fig:fig7}
\end{figure}

\begin{figure}[h]
    \centering
    \includegraphics[width=0.7\linewidth]{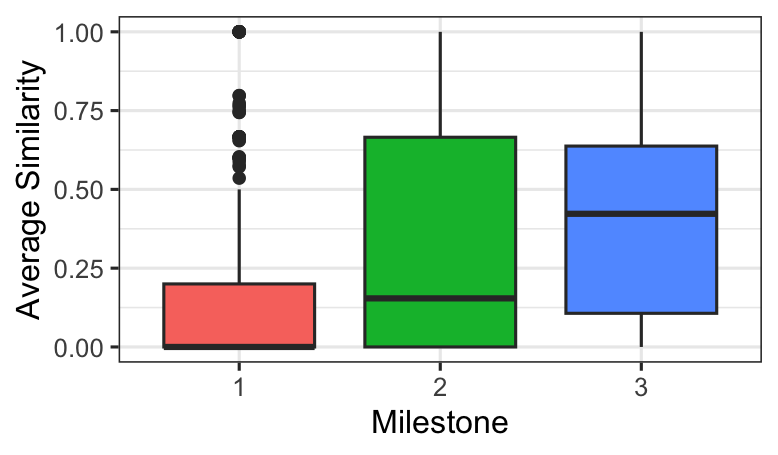}
    \caption{Similarity of generated and integrated code}
    \label{fig:fig9}
\end{figure}
To further assess modification, we compared each instance of repository code with the original ChatGPT code. We calculated the average differences across four metrics: Total Lines of Code, Cyclomatic Complexity, Maximum Control Flow Graph (CFG) Depth, and Halstead Effort. The density plot (Figure \ref{fig:fig7}) shows the distribution of deviations, with a vertical black line marking the point where the difference is zero. The x-axis of the plot was log-transformed for better visualization. The shifted zero point is indicated by the black line, highlighting instances where no change was made to the GPT-generated code. The plot reveals that a significant proportion of deviations are positive, indicating that repository code is frequently more complex than the original GPT version. This trend is consistent across all metrics and milestones, suggesting that students often modify GPT-generated code by increasing its complexity—whether by adding more lines, enhancing logical structures, or deepening control flows. 

To understand modification patterns, we analyzed the logical and structural similarity between repository code and ChatGPT-generated code. We used Jaccard similarity to measure overlap in relevant information extracted via Tree-sitter. Each string pair was assessed with the Longest Common Subsequence (LCS) method, considering pairs over 90\% similar as equivalent. The Jaccard similarity was the ratio of intersection to union of Tree-sitter-extracted sets. Over time, similarity scores highlighted students' evolving use of AI-generated code. As shown in Figure \ref{fig:fig9}, similarity increased across project milestones. In Milestone 1 (MS1), similarity was low and variable, indicating experimentation with AI code. By Milestone 2 (MS2), median similarity rose, suggesting increased reliance on ChatGPT outputs with fewer modifications. At Milestone 3 (MS3), similarity peaked with fewer outliers, reflecting a stronger dependency on generated code.

\begin{chatgroup}{Prompts by teams 5 and 13, during MS1}
    \begin{chatbubble}
    \scriptsize
    After extracting source SIMPLE program into tokens, how do I validate that a cond\_expr is syntactically valid according to the grammar rules?
    \end{chatbubble}
    \begin{chatbubble}
     \scriptsize
    Generate a PKB stub class that I can assign the return value through the constructor. You can work with this:…
    \end{chatbubble}
\end{chatgroup}

An important factor in the increasing similarity scores is the improvement in students' prompting techniques. As they gained experience with ChatGPT, their prompts likely became more refined, leading to more accurate and task-specific outputs. These enhanced outputs could have made the AI-generated code easier to incorporate with minimal changes, contributing to the rising similarity scores. This trend, combined with qualitative observations, has key pedagogical implications. The exploratory behavior in MS1 suggests an active learning phase where students experiment and modify AI-generated solutions, deepening their understanding. As students gained experience, they began using ChatGPT more efficiently, refining prompts to produce high-quality code. By MS3, the workflow stabilized, with consistent similarity scores reflecting a seamless integration of AI into their process.

Chat Listing 2 captures prompts used in MS1, which tend to be straightforward and limited in scope, often yielding basic outputs that require further customization to meet students’ needs.

\begin{chatgroup}{Improved prompting by teams 5 and 13, at the end of the semester in MS2 and MS3}
    \begin{chatbubble}
     \scriptsize
    I need to implement a semantic checker to check that there are no cyclic procedure calls before I build the AST. How should I add on to that for my SemanticValidator class? This is how my SemanticValidator class looks like for now. It receives lines of tokenized code from Tokenizer ..
    \end{chatbubble}
    
    \begin{chatbubble}
     \scriptsize
    Swap the inner sections with the outer ones. E.g., in the section for ‘Contains pair when table is empty,’ have 4 subsections for each combination of $<$int, int$>$, $<$int, string$>$, and so on.
    \end{chatbubble}
\end{chatgroup}

In MS3, students’ prompts become more advanced, specifying examples, constraints, and project-specific contexts. This more strategic prompting enables GPT to produce outputs closely aligned with project requirements, reducing the need for extensive modifications. Thus, the increase in similarity scores over milestones reflects not only students’ growing reliance on GPT but also their refinement in AI interaction, signaling a maturity in prompt engineering that enhances productivity and code quality. 
For educators, this implies the importance of teaching effective prompting techniques and encouraging initial experimentation to ensure that students can critically assess and adapt AI-generated code.

\subsection{In-depth Analysis of Conversations}

\begin{figure}[t]
    \centering
    \includegraphics[width=0.7\linewidth]{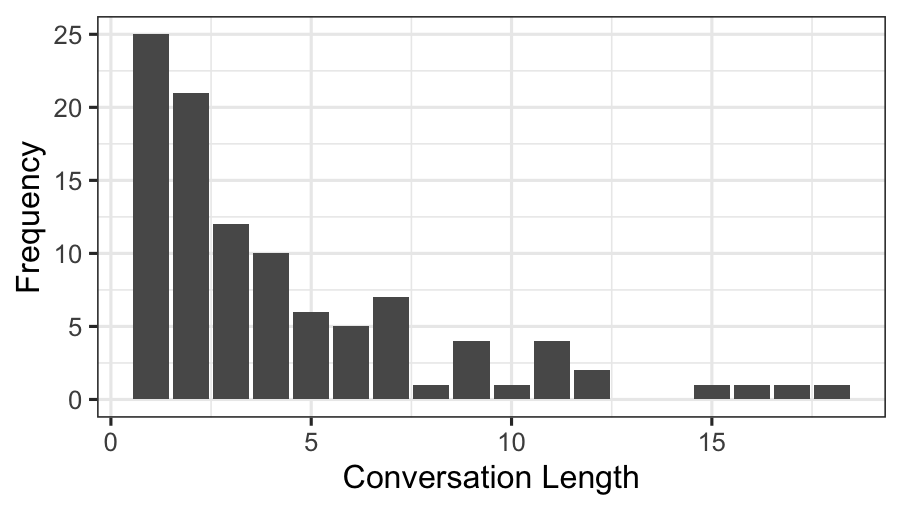}
    \caption{Histogram of Conversation Lengths (filtered for conversations with less than 20 messages)}
    \label{fig:fig12}
\end{figure}

Similarity measurements on the ChatGPT conversations were used to determine how the generated code evolved during a conversation and ultimately integrated. The histogram in Figure \ref{fig:fig12} reveals that most conversations typically consist of just one or two messages, with a smaller number extending beyond 15 messages. The longest conversation was 50 messages. This distribution shows an overall downward trend, indicating that longer conversations are less frequent. 

\begin{figure}[b]
    \centering
    \includegraphics[width=0.7\linewidth]{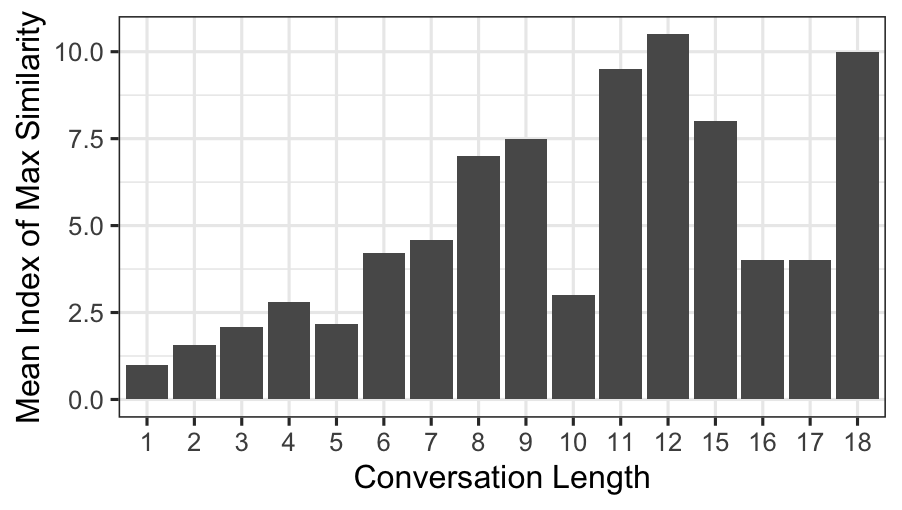}
    \caption{ 
    Mean Index of the Generated Code Most Similar to the Repository Code for Conversations of Different Lengths}
    \label{fig:fig13}
\end{figure}


For each code snippet in the repository, we identified the ChatGPT conversation that generated it by comparing the similarity of GPT-produced code snippets within conversations to the tagged repository code. Conversations were analyzed separately based on varying lengths to account for the tendency of shorter conversations to show high similarity at smaller indices. This separate analysis helped prevent a skew towards smaller indices. For conversations shorter than 20 interactions, we calculated the average index of the code with the highest similarity to the repository code, excluding reused code to avoid skewed values. Conversations averaging zero similarity, suggesting significant modifications or irrelevant outputs, were omitted. In cases of ties in maximum similarity across conversation stages, we prioritized the first occurrence to highlight the initial prompt responsible for the highest similarity.

The results in Figure \ref{fig:fig13} show a general upward trend in the index of the generated code used in the repository as conversations continue. This indicates that as conversations progress, the code ultimately included in the final repository is often generated during the later stages of the dialogue. This trend suggests that students leverage iterative back-and-forth interactions with the LLM to refine and improve the code. However, the mean position of the final code within the conversation is consistently lower than the total conversation length. This implies that the final version of the code does not always originate from the last prompt. Instead, students may opt for earlier outputs that better suit their needs or seek clarification on specific portions of the generated code to enhance their understanding. 

This shows that while LLM-generated code provides valuable starting points, students often interact with the model over several iterations, modifying and adapting the code before integrating it into the final codebase. The increasing index of similarity as conversations progress suggests that students can effectively prompt to make nuanced modifications and refinements to the generated code as required by their use case.

\subsection{Prompt Analysis}
We conducted sentiment analysis for student prompts utilizing the VADER (Valence Aware Dictionary and Sentiment Reasoner) tool \cite{Hutto_Gilbert_2014}. VADER is effective for analyzing the sentiment of short texts, such as prompts, which enables us to determine whether users generally felt positive, neutral, or frustrated during their interactions with the LLM.

\begin{figure}[h]
    \centering
    \includegraphics[width=0.75\linewidth]{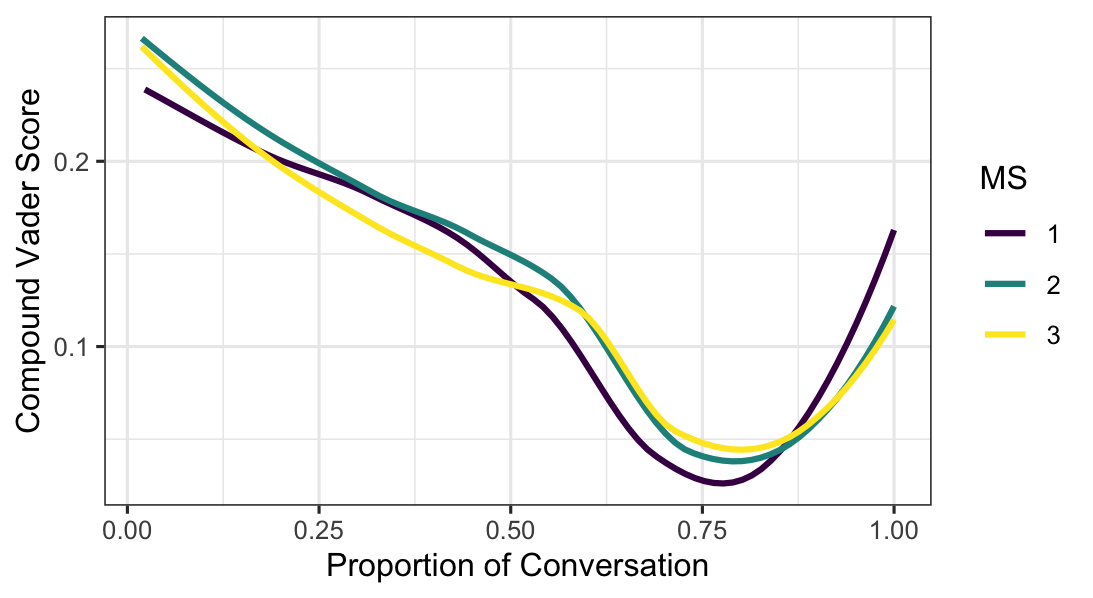}
    \caption{Variation of Compound Vader Scores Over a Conversation: Estimated using LOESS (locally estimated scatterplot smoothing)}
    \label{fig:fig17}
\end{figure}

Figure \ref{fig:fig17} explores sentiment fluctuations throughout individual conversations within each milestone. Initially, the first few messages generally exhibit higher positive sentiment scores, indicating that users often begin interactions with a constructive or hopeful outlook. However, as conversations progress, sentiment tends to show a steady downward trend suggesting that users may encounter challenges or express frustrations as interactions progress, highlighting moments of potential struggle as they clarify questions or seek further assistance. As the conversation nears its conclusion, sentiment stabilizes and ends with a slight uptick of sentiment. This suggests a conclusion to conversations with a sense of resolution.

For example, in one conversation involving unit testing for the AssignParsingStrategy::parse function, the initial message begins optimistically, with the user providing detailed code for context and a clear prompt for assistance. As the conversation progresses, subsequent messages reflect increasing frustration as the user struggles to refine test cases and address specific errors (e.g., “it gives an error saying ‘SyntaxError’ does not refer to a value”). The sentiment recovers slightly toward the end as the issue is resolved, illustrating the characteristic fluctuations in sentiment we observed in many conversations


\section{Threats to Validity}


Our analysis is based on voluntarily collected student self-reports, which may include underreporting or selective disclosure, introducing potential bias. Although GitHub Copilot offers a chat feature, it was not widely used; it primarily served for code completion and debugging. The chat functionality was not significant. Moreover, the VADER tool for sentiment analysis often misclassifies technical terms as neutral, resulting in many prompts receiving scores near zero due to frequent technical language. Despite these limitations, the analysis offers valuable insights into sentiment trends and the emotional tone of user interactions.

\section{Related Work}\label{sec-lit}

\subsection{LLMs in SE Education (LLM4SE Edu)}

The increased popularity and accessibility of LLMs are prompting significant changes to approaches to software engineering education, with an emphasis on adaptive learning strategies and ethical considerations. \cite{3626252.3630927} underscores the need for SE education to evolve in response to LLM advancements, advocating for combining technical skills, ethical awareness, and adaptable learning strategies.  
AI-powered tutors, such as those based on LLMs, have also shown promise in delivering timely and personalized feedback in programming courses.
\cite{savelka2023efficientclassificationstudenthelp} has also found LLMs to be feasible in classifying student needs in SE educational courses, presenting a cost effective alternative to traditional tutor support demand.
However, \cite{3639474.3640061} highlights challenges such as generic responses and potential student dependency on AI, warranting further discussions on the cost-effectiveness of using LLMs in SE education.  
Similarly, \cite{3661167.3661273} finds that Gamified learning environments, when augmented with LLMs, can boost student engagement but may inadvertently lead to over-reliance, undermining the learning process. 
The StudentEval benchmark also introduces novice prompts, shedding light on non-expert interactions and revealing critical insights into user behavior and model performance \cite{llm4code95}.
Work has also been done on programming assistants that do not directly reveal code solutions \cite{10.1145/3613904.3642773}, providing design considerations for future AI education assistants.

\subsection{LLMs in Software Engineering (LLM4SE)}


LLMs have been employed in tools designed to improve code comprehension directly within integrated development environments (IDEs). These tools utilize contextualized, prompt-free interactions to enhance task efficiency, as shown in \cite{3597503.3639187}.  
In the realm of automated unit test generation, ChatGPT has demonstrated competitive performance against traditional tools like Pynguin, particularly when enhanced through prompt engineering techniques \cite{llm4code38}.  
LLMs have also been leveraged to generate insightful questions that bridge gaps between data and corresponding code, improving semantic alignment and comprehension \cite{llm4code21}.  
Automated Program Repair (APR) is another area where LLMs have proven effective, showcasing their ability to fix bugs in both human-written and machine-generated code \cite{10172854}.  
Additionally, \cite{2409.02977} provides a comprehensive survey of LLM-based agents, emphasizing their utility in addressing complex software engineering challenges through human and tool integration.
Despite these promising advancements, \cite{3639476.3639764} highlights critical challenges in ensuring the validity and reproducibility of LLM-based SE research, proposing guidelines to mitigate risks such as data leakage and model opacity.

\subsection{LLMs in Education}
LLMs are promising to reshape pedagogy, by offering solutions for personalized learning and scalable assessment practices. A systematic review of LLM applications in smart education highlights their role in enabling personalized learning pathways, intelligent tutoring systems, and automated educational assessments \cite{2311.13160}. LLMs have also been evaluated for their utility in grading programming assignments, with research demonstrating that ChatGPT provides scalable and consistent grading, rivaling traditional human evaluators \cite{llm4code5}.



Our work extends beyond these existing work in the following aspects: we performed a study on the interaction between LLMs and Software Engineering students working on a complex project, conducting a comprehensive suite of analyses on both the prompts and generated code produced in these interactions, differing from the existing literature in the scope of analysis, a focus on the effects of the conversational nature of LLM code generators, as well as the examination of user sentiments via prompts they used to generate code.

\section{Summary}\label{sec-con}

\textbf{Research Objectives and Contributions: }Our paper explores the integration of Large Language Models (LLMs) in software engineering education, focusing on how student teams interact with AI tools throughout a multi-milestone academic project. We analyzed tool usage, code complexity, refinement, and student prompting behavior to uncover patterns in AI-aided code development throughout the educational process. Our study provides actionable insights for educators to optimize AI tool usage in Software Engineering curricular.

\textbf{Summary of Findings: } Most of the teams utilized AI during development. Copilot was preferred for auto-completion, while ChatGPT excelled in iterative refinement of more complex solutions. AI usage declined across milestones, as students relied on LLMs more at the early stages of the project. Copilot's outputs were often more complex, while ChatGPT produced more concise and understandable solutions. The AI-generated code showed increasing alignment with project goals over time, showcasing improved prompt engineering. Early prompts were exploratory and less precise, later students gained experience and improve on this skill. Sentiment analysis highlighted initial positivity, occasional mid-conversation frustration, and eventual resolution, underscoring the iterative value of AI-assisted coding.

\textbf{Evolution of Student Engagement with LLMs: } Over the course students demonstrated notable growth in their use of LLMs, with improved prompt engineering and more efficient workflows compared to our previous study \cite{Rasnayaka2024}. Access to paid LLMs enabled broader integration of AI tools, encouraging deeper engagement in AI-assisted problem-solving. Increased prevelance of LLM use highlights key pedagogical implications, including the enhanced critical assessment and integration of AI in software development.

\textbf{Implications for Educators: } Experiential learning of Prompt Engineering is effective to enhance code quality and reduce refinement effort. Providing avenues to critically assess AI-generated outputs mitigates over-reliance. Therefore, integration of AI tools in software engineering curricula is essential to maximize the benefits of LLMs. This study highlights the potential of LLMs to transform educational practices, fostering both productivity and deeper understanding in software development processes.

\bibliographystyle{IEEEtran}
\bibliography{references}

\end{document}